\documentclass[11pt, a4paper]{article}

\usepackage{geometry} 
\usepackage{authblk}              
\usepackage[utf8]{inputenc}
\usepackage[compress]{cite}
\usepackage[colorlinks=true, linkcolor=blue, urlcolor=blue, citecolor=blue]{hyperref} 
\usepackage{algorithm}
\usepackage{algpseudocode}
\usepackage{booktabs}
\usepackage{url}
\usepackage[section]{placeins}
\usepackage{hyperref}

\usepackage{amsmath}
\usepackage{amssymb}

\usepackage{graphicx}

\date{}

\title{Simulating two-Dimensional decaying vortex structure with the Carleman-Lattice-Boltzmann method}

\author[1]{Luca Cappelli\thanks{luca.cappelli@iit.it}}
\author[2]{Massimo Bernaschi}
\author[1]{Sauro Succi }

\affil[1]{Fondazione Istituto Italiano di Tecnologia, \protect\\ Center for Life Nano-Neuroscience at la Sapienza, \protect\\ Viale Regina Elena 291, 00161 Roma, Italy}
\affil[2]{CNR, centro per il calcolo ``Marco Picone''}

\begin{document}

\maketitle

\begin{abstract}
We perform a numerical study of the free decay of two-dimensional nonlinear vortex structures
using the second-order-truncated Carleman–Lattice Boltzmann method (CLB2).
To this purpose, we simulate the nonlinear vortex system on a $128\times128$ grid 
at two initial Reynolds numbers, $Re\approx15$ and $Re\approx46$,  by resorting to a 
highly-tuned multi-GPU implementation of the CLB2 code. 

At $Re \approx 15$,  CLB2 shows significant accuracy all along the decay process, up to six 
diffusive times (corresponding to $950$ time steps), with structure-preserving errors below 10\%
in amplitude, where by structure-preserving we imply that such errors can be 
almost entirely reabsorbed by a global time-dependent rescaling factor. 
At $Re \approx 46$, the evolution is carried out over $2850$ time steps,  also corresponding to six diffusive times.  
In this scenario, however, the errors grow in time well above 10\% in amplitude with
a still relatively modest but appreciable effect on the spatial structure of the flow. 

These results indicate that CLB2 is capable of preserving the nonlinear evolution of coherent structures 
over sufficiently long time spans to reach an effective zero steady-state.
However CLB2 shows a constant sub-dissipative trend (more energy than the corresponding
Lattice Boltzmann configurations) with a deteriorating accuracy at increasing time and Reynolds number. 

The present results shed hope that further developments,  such as suitable  closures instead 
of truncation and/or various forms of regularization,  may lead
to a significant improvement of the CLB2 accuracy, prior or even without escalating 
to expensive higher-order truncations.

\end{abstract}

\section{Introduction}

The development of quantum algorithms for fluid dynamics has attracted increasing attention in  recent years~\cite{zylberman2022quantum, meng_clb, demirdjan,Zamora2026, Tennie2025, sanavio_CLB, BharadwajSreenivasan2025, SREENI, bravyi2026, Cappelli26}. 
One of the main motivations is the favourable memory scaling offered by quantum state representations, which could become particularly relevant for large-scale turbulent flow simulations. 
In classical computing, increasing the spatial resolution of a turbulent flow quickly leads to very large computational and memory requirements. This raises the question of whether quantum computing could, in the future, make it possible to simulate turbulent flows at large scales and low viscosities, thereby reaching higher Reynolds numbers.  
There is, however, a fundamental difficulty in translating fluid dynamics into a quantum framework: quantum evolution 
is linear and conservative, whereas the equations governing fluid motion are inherently nonlinear and dissipative~\cite{NielsenChuang2000, Succi_2026_EPL}.

Among the possible strategies to overcome these difficulties, the Carleman linearization is one of the most promising~\cite{Carleman}. The idea is to embed the non-linear system into a higher-dimensional one, by introducing new variables corresponding to the nonlinear terms. The resulting system is formally infinite, and in practical calculations it has to be truncated at some order \(K\), neglecting terms of higher order. This approach is particularly interesting when applied to Lattice Boltzmann models, where the structure of the equations and their local nature provide a convenient starting point for the Carleman formulation~\cite{Itani2022, Turro_25, Sanavio_25_PoF, wiebe25}.

However, Carleman linearization does not guaranty convergence to the exact solution of the 
original nonlinear system~\cite{Amini2025, forets2017}. 
At high Reynolds numbers, Carleman truncation errors may increase with truncation order~\cite{psiq_26}, while a related runaway behaviour is suggested in~\cite{cappelli_bernaschi_succi_26}, potentially limiting the time interval over which convergence can be maintained.
In the latter situation, the linearized system may diverge from the exact solution regardless of the finite order $K$ considered. 
This behavior is typical of the logistic equation in its growing regime and can also occur when forcing 
induces local instabilities. 
Hence, it appears that an {\it all-time} convergence of the Carleman procedure
at sufficiently high Reynolds number might be very hard to achieve.
On the other hand, many practical fluid problems do not demand detailed information
all along the time evolution, but only at (statistical) steady state.
This is a much less stringent requirement, and it has indeed previously been shown that 
even the lowest-order CLB2 approximation can recover the steady state of low-Reynolds flows~\cite{Cappelli_PoF, cappelli_bernaschi_succi_26}.
\\
Several theoretical bounds have been established for simpler fluid systems, such as the Burgers equation~\cite{Liu2021, Conde2025, gonzalezconde2026stability, Jennings2026quantumalgorithms}. 
These results typically aim to guarantee exponential or uniform convergence to the exact solution. However, to the best of our knowledge, the behavior of Carleman linearization for turbulent-like flows has not been investigated yet, due to the high computational cost of the Carleman procedure.

As a first step in this direction, we investigate whether the lowest-order Carleman–Lattice Boltzmann (CLB2) approximation can reproduce the decay of two-dimensional coherent structure (vortices) configurations. 
We also aim to identify the limits of the method and determine when second-order Carleman truncation fails
to capture the system dynamics accurately. To this end, we use a multi-GPU implementation of CLB2 truncated at second order. With the current code and computational resources, we can simulate the system on a $128\times128$ grid on
a moderately large GPU-cluster platform.

The rest of the paper is organized as follows. In Sec.~\ref{sec:CLB}, we introduce the Lattice Boltzmann framework and its Carleman linearization for unforced evolution. We then briefly introduce the multi-GPU code we employed for the simulation in Sec.~\ref{sec:Numerical resutls}.
Subsequently, in Subsec.~\ref{subsec:initial_conditions} we present the initial conditions used to generate turbulent-like behavior and the metrics used to compare CLB2 with the reference solution. Finally, we present results for two cases with the same initial condition but different viscosities. 
When the viscous effects increase, the second-order-truncated Carleman (CLB2) correctly reproduces the nonlinear decay of the vortices and captures the energy spectrum while preserving the non-zero modes. 
However, as the viscosity decreases, the second-order approximation is no longer sufficient to accurately capture the turbulent-like decay.

\section{Carleman Lattice Boltzmann}\label{sec:CLB}
Before discussing the numerical implementation of Carleman lattice Boltzmann, we briefly recall the main ingredients of the lattice Boltzmann method and its second-order Carleman formulation.
\\
Consider the lattice Boltzmann Method (LBM), specifically a forced D2Q9 model~\cite{Succi2001, BENZI1992}, consisting of nine discrete distributions in two spatial dimensions. The fluid is described by a set of distribution functions, or populations, $f_i$, each associated with a discrete lattice velocity $\mathbf{c}_i=(c_{xi},c_{yi})$. In the D2Q9 model, the velocity components take the values $-1$, $0$, or $1$, giving a total of $Q=9$ populations.
\\
The dynamics is divided into two steps: collision and advection. The collision, or relaxation, step is local but nonlinear and is given by
\begin{equation}
\label{eq:collision}
f^*_i(\mathbf{x},t)
=
(1-\omega)\,f_i(\mathbf{x},t)
+\omega\,f_i^{\mathrm{eq}}(\mathbf{x},t)
+F_i(\mathbf{x}) \,,
\end{equation}
where $F_i(\mathbf{x})$ is an external forcing term, $f_i^{\mathrm{eq}}(\mathbf{x},t)$ is the equilibrium distribution, and $\omega$ is the relaxation frequency parameter. The latter is related to the kinematic viscosity through
\begin{equation}
\label{eq:viscosity}
\nu
=
c_s^2\left(\frac{1}{\omega}-\frac{1}{2}\right),
\end{equation}
with $c_s$ denoting the lattice speed of sound. The equilibrium distribution is
\begin{equation}
\label{eq:equilibrium}
f_i^{\mathrm{eq}}(\mathbf{x},t)
=
w_i\rho(\mathbf{x},t)
\left(
1
+\frac{\mathbf{u}\cdot\mathbf{c}_i}{c_s^2}
+\frac{(\mathbf{u}\cdot\mathbf{c}_i)^2}{2c_s^4}
-\frac{\mathbf{u}\cdot\mathbf{u}}{2c_s^2}
\right),
\end{equation}
where $w_i$ are the lattice weights and $\mathbf{u}$ is the local fluid velocity.
\\
The advection, or streaming, step is instead linear and nonlocal,
\begin{equation}
\label{eq:streaming}
f_i(\mathbf{x}+\mathbf{c}_i\Delta t,t+\Delta t)
=
f_i^*(\mathbf{x},t).
\end{equation}
\\
The macroscopic density and current density are obtained from the zeroth- and first-order moments of the populations,
\begin{equation}
\label{eq:observables}
\rho = \sum_{i=1}^Q f_i,
\qquad
\mathbf{J} = \sum_{i=1}^Q f_i \,\mathbf{c}_i = \rho\,\mathbf{u}.
\end{equation}
Substituting these relations into Eq.~\eqref{eq:equilibrium}, the collision step can be expressed directly in terms of the populations. For incompressible or weakly compressible flows, characterized by a small Mach number, $Ma=|\mathbf{u}|/c_s\ll1$, it can be written in the generic polynomial form
\begin{equation}
\label{eq:tensor_collision}
f^*_i
=
A_{ij} \, f_j
+B_{ijk} \, f_jf_k
+C_{ijkl} \, f_jf_kf_l \, .
\end{equation}
The cubic term originates from the approximation $1/\rho\approx2-\rho$, which is valid for $\rho\approx1$ in the weakly compressible regime. The explicit expressions of the tensors $A$, $B$, and $C$ can be found in~\cite{sanavio_CLB}. For compactness, the space-time dependence is omitted since the collision step is local.
\\
This polynomial form makes the nonlinear structure of the collision operator explicit and serves as the starting point for the Carleman linearization.
\\
We can now introduce the Carleman linearization of the LBM dynamics. Defining the second-order lifted variables
\begin{equation}
\label{eq:carleman_variables}
g_{ij}(\mathbf{x}_1,\mathbf{x}_2)
=
f_i(\mathbf{x}_1)f_j(\mathbf{x}_2),
\end{equation}
the nonlinear collision operator is mapped onto a hierarchy involving products of increasingly high order. Truncating this hierarchy at second order yields the Carleman lattice Boltzmann (CLB) approximation, denoted by CLB2. In compact notation, the resulting collision step takes the form
\begin{align}
\label{eq:C1}
f_i^* &= A_{ij}\,f_j + B_{ijk}\,g_{jk},
\\
\label{eq:CLB2}
g_{ij}^* &= A_{ik}A_{jl}\, g_{kl}.
\end{align}

The dimensions of the tensors $A$ and $B$ are not, by themselves, the main source of computational difficulty in the implementation. Rather, the primary bottleneck arises from the non-local nature of the second-order Carleman tensor in Eq.~\eqref{eq:carleman_variables}. In fact, for a grid of size $G$ and a Carleman truncation order $K$, the total number of variables scales as $(QG)^K$. For $Q=9$ and a $128^2$ grid, for example, a second-order truncation requires $81\times128^4$ elements, corresponding to approximately $162\,\mathrm{GByte}$ in double precision.

The non-locality of $g_{ij}(\mathbf{x}_1, \mathbf{x}_2)$ is, in turn, a direct consequence of the non-locality introduced by the streaming operation. The resulting memory requirements make the simulation of large systems on conventional computing architectures challenging and call for efficient strategies to limit data movement and memory consumption. In the following section, we therefore describe the multi-GPU implementation adopted in this work.

\section{Numerical Results}\label{sec:Numerical resutls}
As discussed in the previous Section, directly simulating a Carleman-linearized system requires substantial memory and computational time. Running these simulations on a $128\times128$ grid therefore requires a dedicated implementation. For this reason, we developed a multi-GPU code to handle the large memory requirements.
\\
\subsection{Code implementation }\label{subsec:code}
At its core, the code divides the computational grid into smaller spatial regions, each assigned to a different GPU. At every time step, each GPU updates its local region and exchanges the boundary data needed by neighboring GPUs to propagate the solution across the entire domain. To improve efficiency, computations on interior grid points proceed while this data exchange is in progress. To this purpose, asynchronous memory copies and non-blocking message passing operations are employed. In addition, since the Carleman formulation requires pairwise information spanning the entire domain, the corresponding global field is synchronized across GPUs through collective communication. By distributing the computational workload and overlapping communication with computation, this implementation makes it possible to simulate the memory-intensive Carleman system on a \(128\times128\) grid.
\\
The efficient multi-GPU architecture allowed us to simulate the evolution of a coherent system on a $G=128\times128$ grid.
While the code would be able to implement even the forced scenario,
we have chosen to focus on the free-decay case as this represents our first attempt to investigate vortical systems in a two-dimensional setting.
\\
In this Section, we detail how the initial condition was prepared and present the results obtained from both a physical and an HPC perspective.

\subsection{Initial condition}\label{subsec:initial_conditions}
The simulations start from a homogeneous and isotropic velocity field with
zero mean. The initial condition is generated in Fourier space by assigning
to each independent Fourier mode a complex amplitude
\begin{equation}
a(\mathbf{n}) =
\sqrt{S(n)}\,e^{i\phi_{\mathbf{n}}},
\qquad
S(n)=
\left(\frac{n}{n_0}\right)^4
\exp\left[-2\left(\frac{n}{n_0}\right)^2\right],
\label{eq:initial_spectrum}
\end{equation}
where $\phi_{\mathbf{n}}$ is a random phase uniformly distributed in
$[0,2\pi)$ 
and $n$ labels the discrete Fourier modes of a periodic square domain of linear size $L$
\begin{equation}
\mathbf{k}=\frac{2\pi}{L}\mathbf{n},
\qquad
k=|\mathbf{k}|=\frac{2\pi}{L}n.
\end{equation}
The distribution is thus centered around the characteristic mode
$n_0$ and excites a range of neighboring scales, providing a vortical
initial state for the subsequent nonlinear decay of the fluid flow.
In the present simulations we used $n_0=4$ as a default parameter.

The velocity Fourier modes are constructed in the direction perpendicular
to the mode vector,
\begin{equation}
\hat{\boldsymbol{u}}(\mathbf{n})=
a(\mathbf{n})
\frac{(-n_y,n_x)}{n},
\qquad n\neq0,
\label{eq:initial_velocity}
\end{equation}
so that
\begin{equation}
\mathbf{n}\cdot
\hat{\boldsymbol{u}}(\mathbf{n})=0
\end{equation}
for every mode. The initial velocity field is therefore divergence-free
by construction. The real-space field is then obtained through an inverse
Fourier transform.

Finally, 
the density is initialized uniformly with $\rho=1$ and the initial distribution $f_i$ is obtained by inserting the velocity field in the equilibrium expression in Eq.~\eqref{eq:equilibrium}.

\subsection{Numerical results}
Our primary goal was to assess whether CLB2 could accurately reproduce the 
free decay of a two-dimensional turbulent-like flow. 
To this end, we first simulated a system with moderate viscosity, $\nu=1/6$, on a $128\times128$ grid, setting $\omega=1$.

Using the following definition of the Reynolds number,
\begin{equation}\label{eq:reynolds}
Re(t)=\frac{U_{\mathrm{RMS}}(t) \,N}{\nu}\,, \qquad 
U_{\mathrm{RMS}}(t) = \sqrt{
\frac{1}{G}\sum_{ij}
\left(u_x(x_i, y_j)-\langle u_x\rangle\right)^2 
+ 
\left(u_y(x_i, y_j)-\langle u_y\rangle\right)^2
}\,,
\end{equation}
we obtain $Re(0)\approx15$ for the configuration described above.
Since the dynamics is decaying,  the maximum value is the initial one.
With this initial Reynolds number, the Carleman-linearized system is expected to reproduce the evolution of the system correctly.  We nevertheless aim to assess its accuracy in more detail, focusing in particular on its ability to reproduce the nonlinear transient,  since the zero steady state is attained in any case.

\subsubsection{Initial Reynolds $Re(0)\approx15$}

We evolved the initial condition described in Sec.~\ref{subsec:initial_conditions} for $950$ time steps, corresponding to six viscous times, $T_\nu=1/(\nu k_0^2)$, where $k_0=8\pi/N$ is the dominant mode in the initial configuration. This normalization provides a dimensionless measure of time relative to the viscous decay timescale of the initial dominant mode.

\begin{figure}
    \centering
    \includegraphics[width=\linewidth]{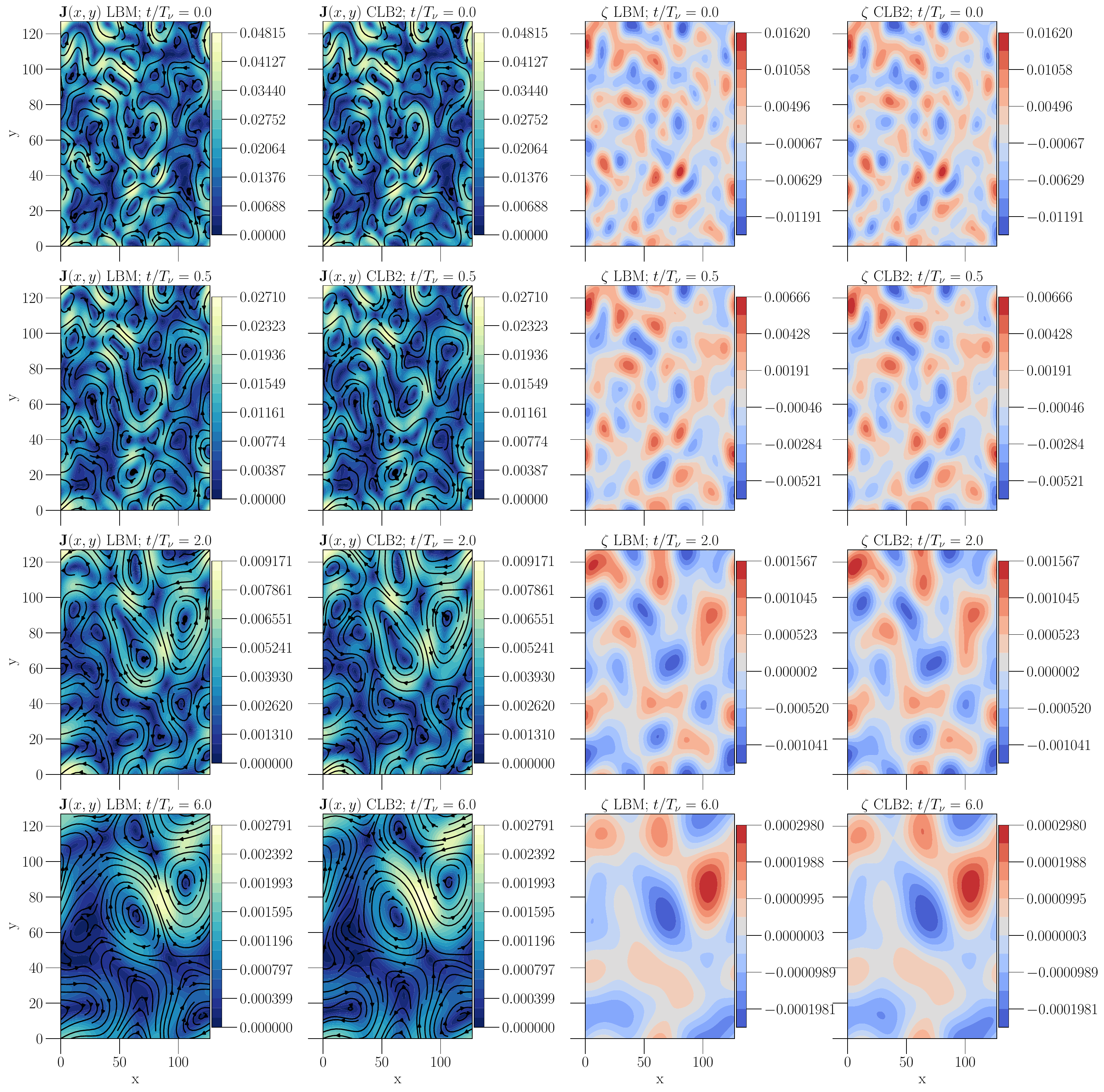}
    \caption{
    Evolution of the momentum field $\mathbf{J}$ and vorticity $\zeta$ for a turbulent-like decay at $Re(0)\approx15$, simulated with Lattice Boltzmann and Carleman truncated at second order (CLB2) on a $128\times128$ grid with $\omega=1$.
    \\
    The first two columns show the magnitude of the momentum field $\mathbf{J}$, together with its streamlines, for LBM and CLB2, respectively, whereas the last two columns show the vorticity obtained with LBM and CLB2, respectively. The reported time frames are expressed in units of the initial characteristic viscous timescale, $T_\nu=1/(\nu k_0^2)\approx156$ time steps.
    \\
    The figure shows that CLB2 closely reproduces the nonlinear decay observed with LBM: the vortices remain in approximately the same positions, and their amplitudes are also in good agreement.
    }
    \label{fig:fields_Re_15}
\end{figure}

Specifically, Fig.~\ref{fig:fields_Re_15} shows the momentum field $\mathbf{J}$, along with its streamlines, and the vorticity $\zeta=\partial_x u_y-\partial_y u_x$ at different time frames. At $t=0$, the initial turbulent-like configuration exhibits small-scale vortical structures. 

At $t/T_\nu\approx1/2$, during the nonlinear transient, the dominant initial mode is expected to have lost
a significant fraction of its initial amplitude due to linear dissipation. 
 At the same time,  before being depleted by dissipative decay,
 the smaller vortices begin to merge into larger ones,  as clearly visible 
 from the map of the vorticity field,  thereby slowing down the exponential decay 
 of the large structures.
This process continues up to $t/T_\nu\approx2$, by which point the initial 
dominant mode is expected to have decayed significantly.  
We extend the simulation to $t/T_\nu\approx6$, when the initial vortices have merged 
into approximately four large vortices.

Throughout this evolution, the momentum and vorticity fields obtained with 
second-order-truncated Carleman (CLB2) are in close visual agreement 
with those obtained with LBM, both during the nonlinear transient and at later times.

This is due to the fact that nonlinear effects become less and less dominant as
the decay proceeds, with the initial Reynolds $Re(0)\approx15$ decreasing to
$Re(T_\nu/2) \approx8$, down to $Re(2T_\nu) \approx3$ and $Re(6 T_\nu) \approx 0.8$.

\begin{figure}
    \centering
    \includegraphics[width=.7\linewidth]{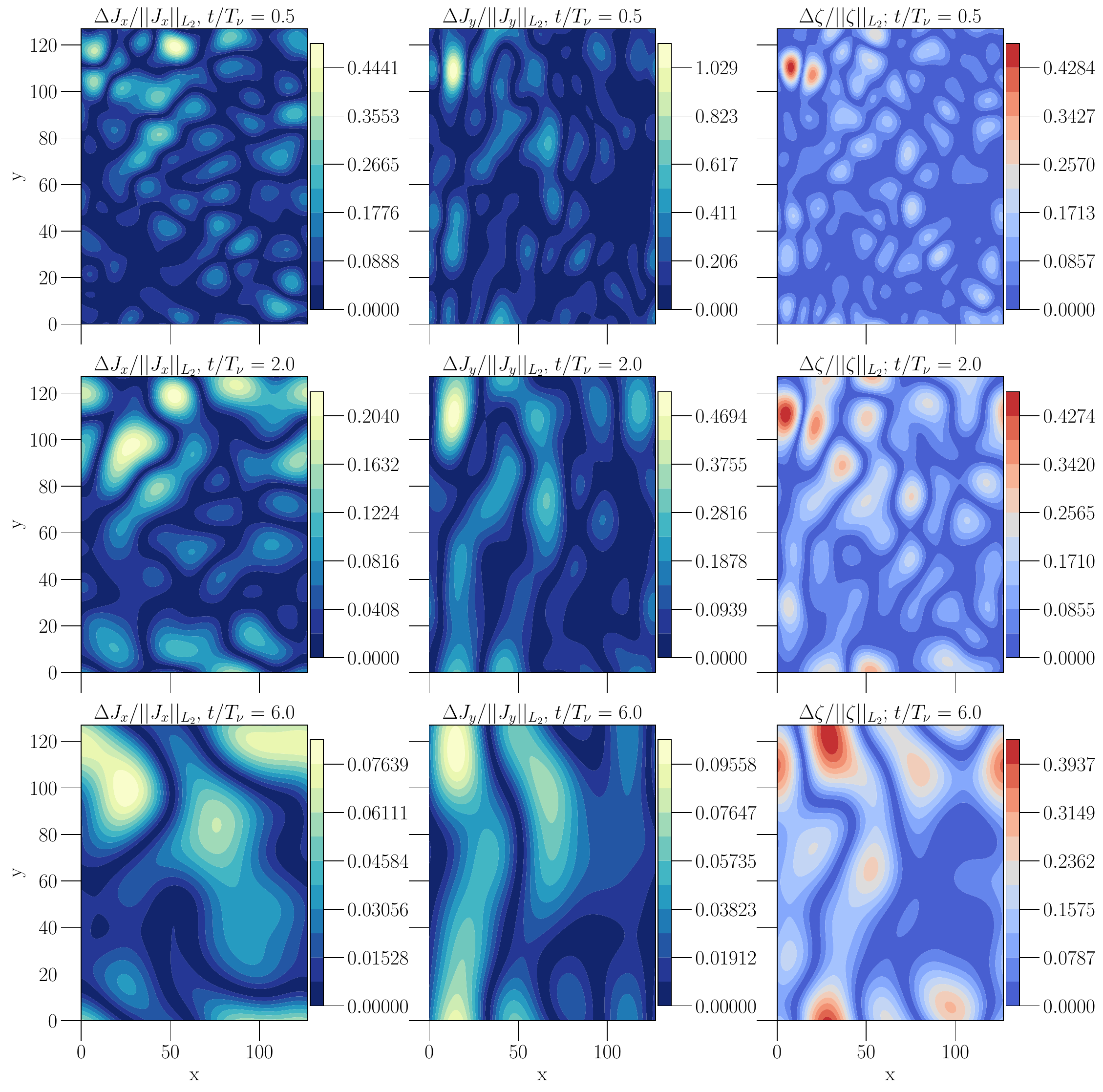}
    \caption{
    Evolution of the relative error in the two components of the momentum field $\mathbf{J}$ and in the vorticity $\zeta$. The figure refers to the scenario shown in Fig.~\ref{fig:fields_Re_15}, with initial Reynolds $15$. The error is evaluated using Eq.~\eqref{eq:error}, and the time frames are normalized with respect to the characteristic viscous timescale $T_\nu\approx156$. The first two columns show the errors in the two components of the momentum field, whereas the rightmost column shows the error in the vorticity $\zeta$.
    \\
    Despite some local peaks, the relative error remains below $10\%$ in the majority of the spatial domain throughout the evolution, indicating good agreement between LBM and CLB2. However, the global error appears to increase slightly as the evolution progresses. 
    }
    \label{fig:errors_re_15}
\end{figure}

 
To quantify the agreement between the two methods,  we reported in Fig.~\ref{fig:errors_re_15} the relative errors in the two components of the momentum field and in the vorticity at the same time frames. As an example we report the explicit relation we used for the vorticity discrepancy:
\begin{equation}
    \label{eq:error}
    \frac{\Delta \zeta}{\zeta}
    \big( x_i,y_i \big)
    =
    \frac{\left|\zeta_{CLB2}(x_i,y_i) - \zeta_{LBM}(x_i,y_i)\right|}
    {\sqrt{
    \sum_{ij} \zeta_{LBM}(x_i, y_j)^2/ G
    }} \, .
\end{equation}
 Since the fields decay towards zero, we normalize their difference by a characteristic field magnitude evaluated using the $L_2$ norm, rather than by the local field value. 
This avoids artificially large relative errors at points where the fields are 
close to zero, while still providing a measure  of the discrepancy between the two solutions.
 \\
The local error reaches a peak during the transient ($t=T_\nu/2$), mostly confined to 
the upper-left region of the domain. Elsewhere, the error remains below $10\%$ over most 
of the domain. As the system evolves, this peak gradually disappears, while the regions with larger errors expand, suggesting an increase in the overall error. Nevertheless, the error remains below $10\%$ over most of the domain throughout the evolution, consistent with the qualitative agreement observed in Fig.~\ref{fig:fields_Re_15}.

\begin{figure}
    \centering
    \includegraphics[width=.7\linewidth]{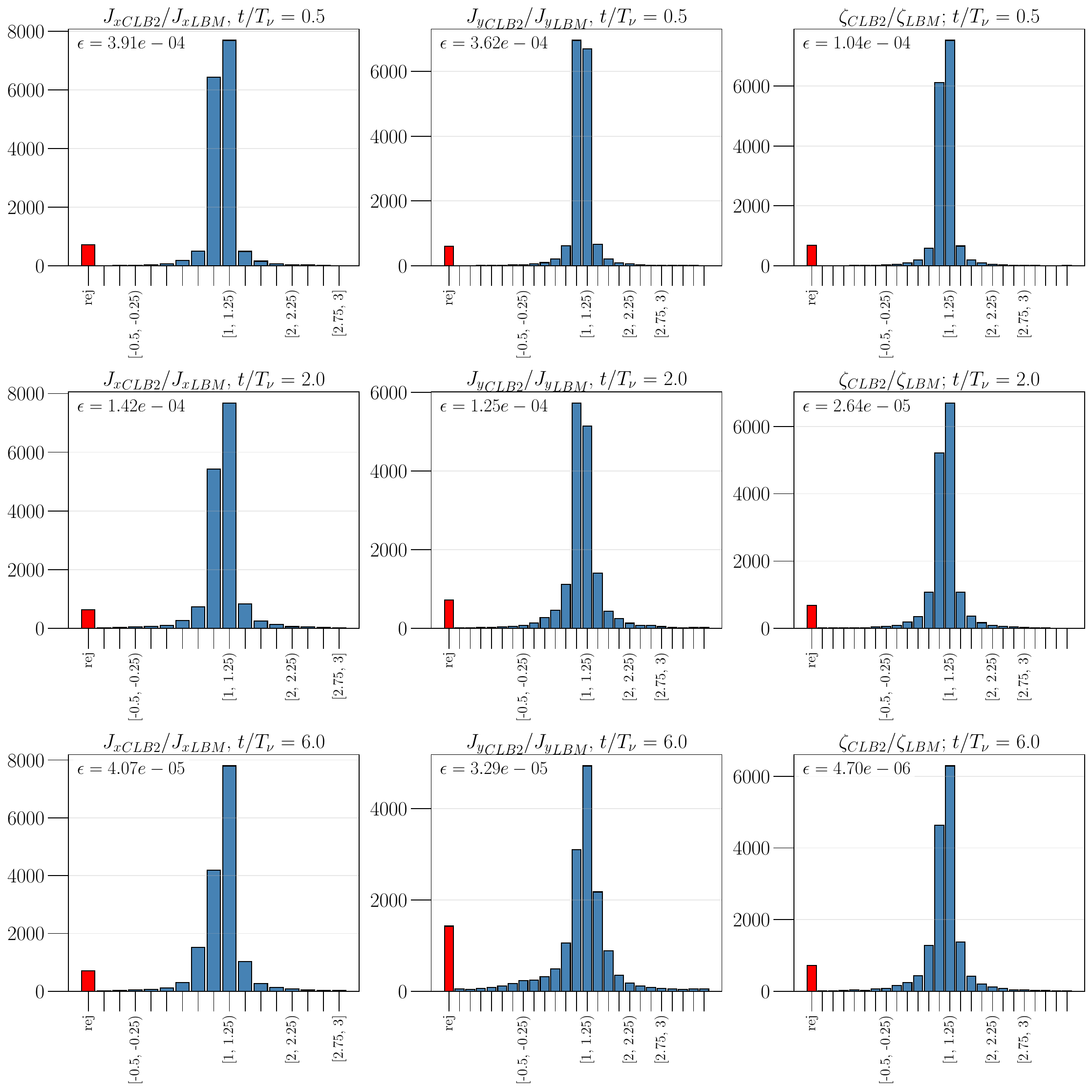}
    \caption{
   Histogram of the ratio between the fields obtained with LBM and CLB2 for the case in Fig.~\ref{fig:fields_Re_15} with initial Reynolds $15$. The histogram shows the number of grid points $(x_i,y_j)$ for which the ratio between the CLB2 and LBM fields falls within each bin along the horizontal axis. The column layout and selected time frames are the same as in Fig.~\ref{fig:errors_re_15}. The red bin labelled $rej$ collects the values excluded from the histogram, either because the corresponding field values fall below the threshold indicated in the upper-left corner or because the ratios lie outside the range covered by the bins. 
   The threshold is evaluated according to Eq.~\eqref{eq:threshold}.
    \\
    A distribution sharply peaked around $1$, together with a negligible number of rejected values, indicates a good agreement between the two solutions, as observed in this case.
    }
    \label{fig:hsito_re_15}
\end{figure}

To further assess the preservation of the structural features of the flow, 
in Fig.~\ref{fig:hsito_re_15} we show the spatial distribution of the ratios 
between the CLB2 and LBM fields at the time frames considered above. 
To avoid anomalously large ratios caused by very small values of the reference 
field, we applied a mask to the LBM field, excluding values below a given threshold. 
Specifically, we excluded points where $|\zeta_{\mathrm{LBM}}(x_i,y_j)|<\epsilon$, with the threshold defined in terms of the RMS of the LBM field as
\begin{equation} \label{eq:threshold}
    \epsilon = 0.05\,\sqrt{\frac{1}{G}\sum_{ij}\zeta_{\mathrm{LBM}}(x_i,y_j)^2}\,.
\end{equation}
We then counted the CLB2/LBM ratios falling within each bin. The bins are spaced at intervals of $0.25$ and cover the range around $1$, extending to the first positive and negative integer values. As expected, the histogram shows very good agreement between LBM and CLB2, with the $\mathrm{CLB2}/\mathrm{LBM}$ ratio distribution sharply peaked around $1$.


Next, we investigate the evolution of the energy spectrum. A characteristic feature of two-dimensional turbulence is the coalescence of vortices into larger structures, accompanied by an inverse energy cascade that transfers energy from smaller to larger scales (i.e., from higher to lower Fourier wavenumbers).
 
\begin{figure}
    \centering
    \includegraphics[width=\linewidth]{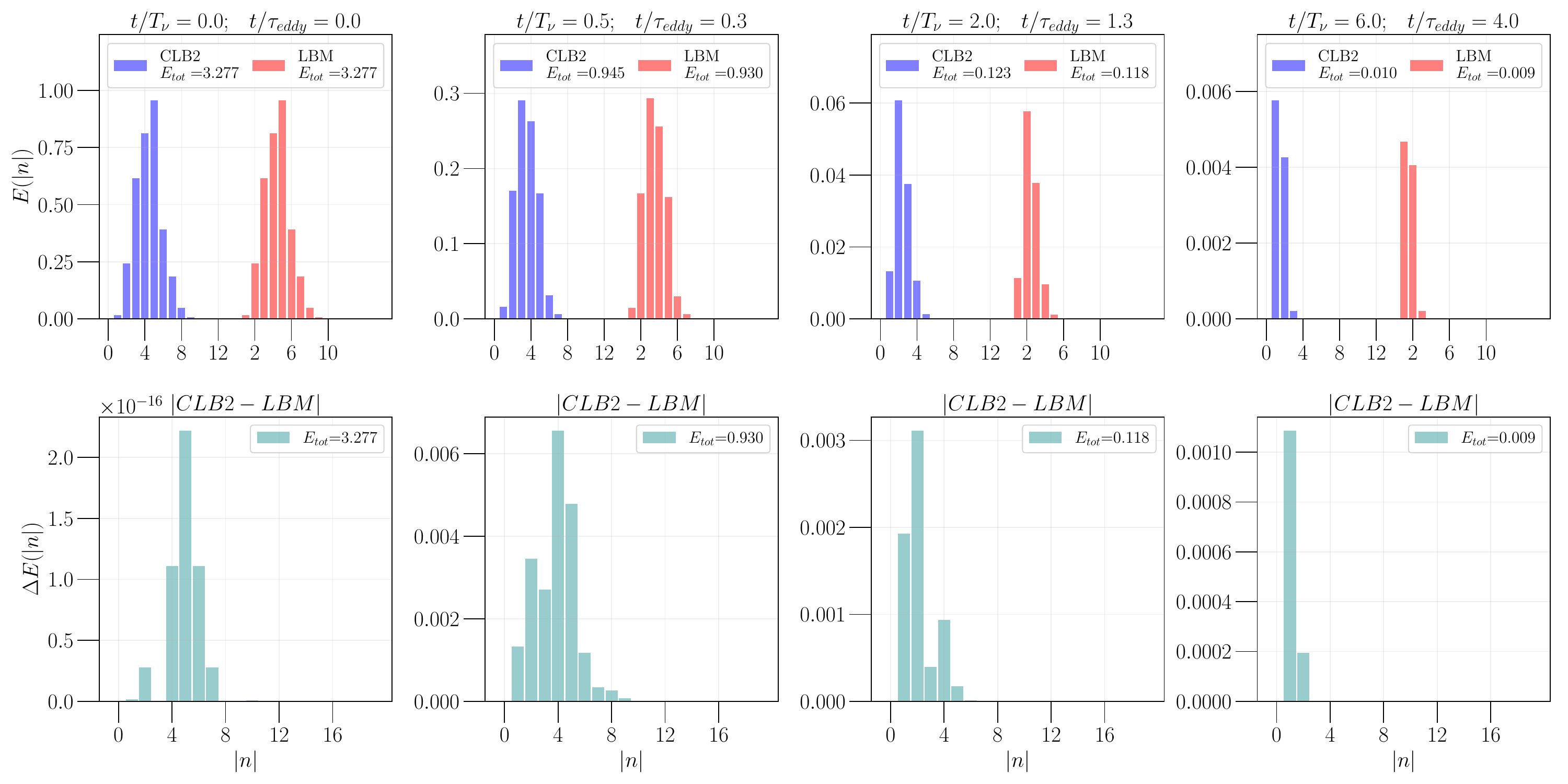}
    \caption{
    (Top row) Comparison of the energy spectra obtained with LBM and CLB2 at the same time frames as in Fig.~\ref{fig:fields_Re_15}. For each time frame, the reported time is normalized with respect to both the characteristic viscous time scale and the eddy turnover time of the initial configuration $\tau_{\mathrm{eddy}}$. The $x$-axis shows the integer modes $|n|$, while the $y$-axis shows the energy contained in each spectral shell, between $|n-1|$ and $|n|$. The total energy at each time frame is reported in the corresponding panel, highlighting its decay over time. Accordingly, the $y$-axis scale differs between panels.
    \\
    (Bottom row) Evolution of the absolute errors for each mode at the same time frames shown in the top row. 
    The figure shows that CLB2 correctly reproduces the energy spectrum of the non-zero modes during the evolution without losing information. The main source of error is the slightly higher total energy obtained with CLB2 than with LBM, as reported in the panel captions.
    }
    \label{fig:energy_re_15}
\end{figure}

We report the radial energy spectrum at the time frames considered above in Fig.~\ref{fig:energy_re_15}. The top row shows the energy associated with each mode for both CLB2 and LBM, together with the total energy.

In the same figure, we also report time in units of the eddy-turnover time, $\tau_{\mathrm{eddy}}$, defined as
\begin{equation}
    \tau_{\mathrm{eddy}}=\frac{\ell_E}{U_{\mathrm{rms}}}\,,
\end{equation}
where $\ell_E$ is the characteristic scale of the energy-containing eddies and $U_{\mathrm{rms}}$ is the root-mean-square velocity fluctuation. 
The characteristic eddy scale \(\ell_E\) is calculated from the two-dimensional Fourier transform of the velocity fluctuations. It is defined as an energy-weighted average of the inverse wavenumber,
\begin{equation}
\ell_E(t)=
\frac{
\sum_{\mathbf{k}\neq 0}
E_{\mathbf{k}}(t) / |\mathbf{k}|
}
{
\sum_{\mathbf{k}\neq 0}E_{\mathbf{k}}(t)
},
\qquad
E_{\mathbf{k}}(t)=\frac{1}{2}
\left(|\hat u_x(\mathbf{k},t)|^2+
|\hat u_y(\mathbf{k},t)|^2\right).
\end{equation}
This definition gives greater weight to large-scale structures, associated with low 
wavenumbers, and provides a characteristic length scale of the energy-containing eddies. 
The associated timescale characterizes the lifetime of the coherent structures before dissipation takes over. 

 The labels in Fig.~\ref{fig:energy_re_15} highlight the effect of viscosity, which reduces the total energy from approximately $3.3$ to $5\times10^{-3}$, as well as the transfer of energy toward lower modes during the decay. The spectrum initially extends up to mode $8$. By $t/T_\nu\approx2$, only the first four modes remain visible, and at the final time only two modes remain. Throughout this process, CLB2 generally yields higher energy values than LBM, but still reproduces the overall spectral distribution well during the nonlinear transient. Small discrepancies become visible at $t/T_\nu\approx6$. At this point, however, the total energy is already close to zero, making the relative error less informative.

Overall, these results show that CLB2 correctly reproduces the turbulent-like decay at $Re(0)\approx15$, capturing both the initial evolution of the fields and the transfer of energy towards larger scales.

\subsubsection{Initial Reynolds $Re(0)\approx46$}

Next,  we reduce the viscosity to $\nu=1/18$, setting $\omega=3/2$,  corresponding to an
initial Reynolds number of approximately $46$. 
Using the same initial condition as in the previous case, we evolve the system over six viscous times, $T_\nu$, corresponding to a total of $2850$ time steps, and repeat the same analysis as
for the case  $Re(0)\approx15$.
In this scenario, the Reynolds number decreases from its initial value to 
$Re(T_\nu/2)\approx24$, $Re(2T_\nu)\approx9$, and finally $Re(6T_\nu)\approx3$. 
Thus, even in this case, nonlinear effects become progressively weaker as the 
system evolves, although constantly above the laminar value $Re=1$.

\begin{figure}
    \centering
    \includegraphics[width=\linewidth]{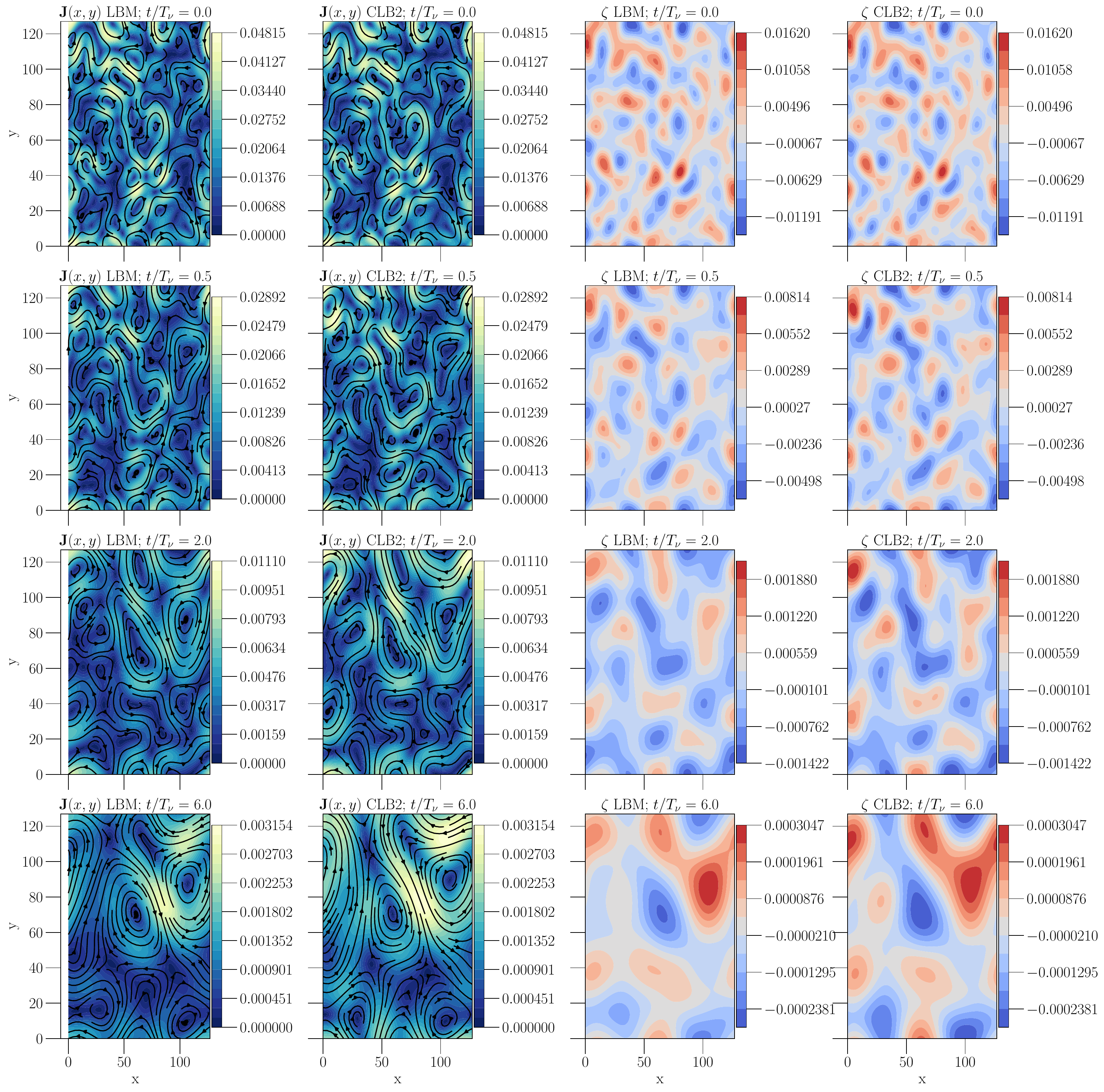}
    \caption{
    Evolution of the momentum field $\mathbf{J}$ and vorticity $\zeta$ during a turbulence-like decay at $Re(0)\approx46$, simulated with Lattice Boltzmann and second-order-truncated Carleman CLB2) on a $128\times128$ grid with $\omega=3/2$. The figure layout is the same as in Fig.~\ref{fig:fields_Re_15}, and the characteristic viscous timescale $T_\nu$ corresponds to approximately $467$ time steps.
    \\
    The same color scales are used for LBM and CLB2 to make even small differences between the results visible. In general, CLB2 produces sharper structures and larger regions of high amplitude at $t/T_\nu\approx6$.
    \\
    As in the previous case, small vortices merge into larger vortical structures as the flow decays.
    }
    \label{fig:fields_re_46}
\end{figure}

Figure~\ref{fig:fields_re_46} shows the evolution of the momentum field $\mathbf{J}$ and the vorticity $\zeta$. 
During the transient, at $t/T_\nu=1/2$, no significant differences between CLB2 and LBM are visible. 
At later times, however, the discrepancies become more apparent. More specifically, CLB2 exhibits regions where the momentum field $\mathbf{J}$ has larger amplitudes than the corresponding LBM field. The streamlines also follow slightly different directions, and the vortices tend to exhibit larger amplitudes and spatial extents. Nevertheless, the overall shape of the fields remains reasonably well aligned. 
These observations suggest that the main differences concern the amplitudes and scales of the structures, rather than the overall flow configuration. To better understand the nature of these discrepancies, we now turn to a quantitative analysis.

\begin{figure}
    \centering
    \includegraphics[width=.7\linewidth]{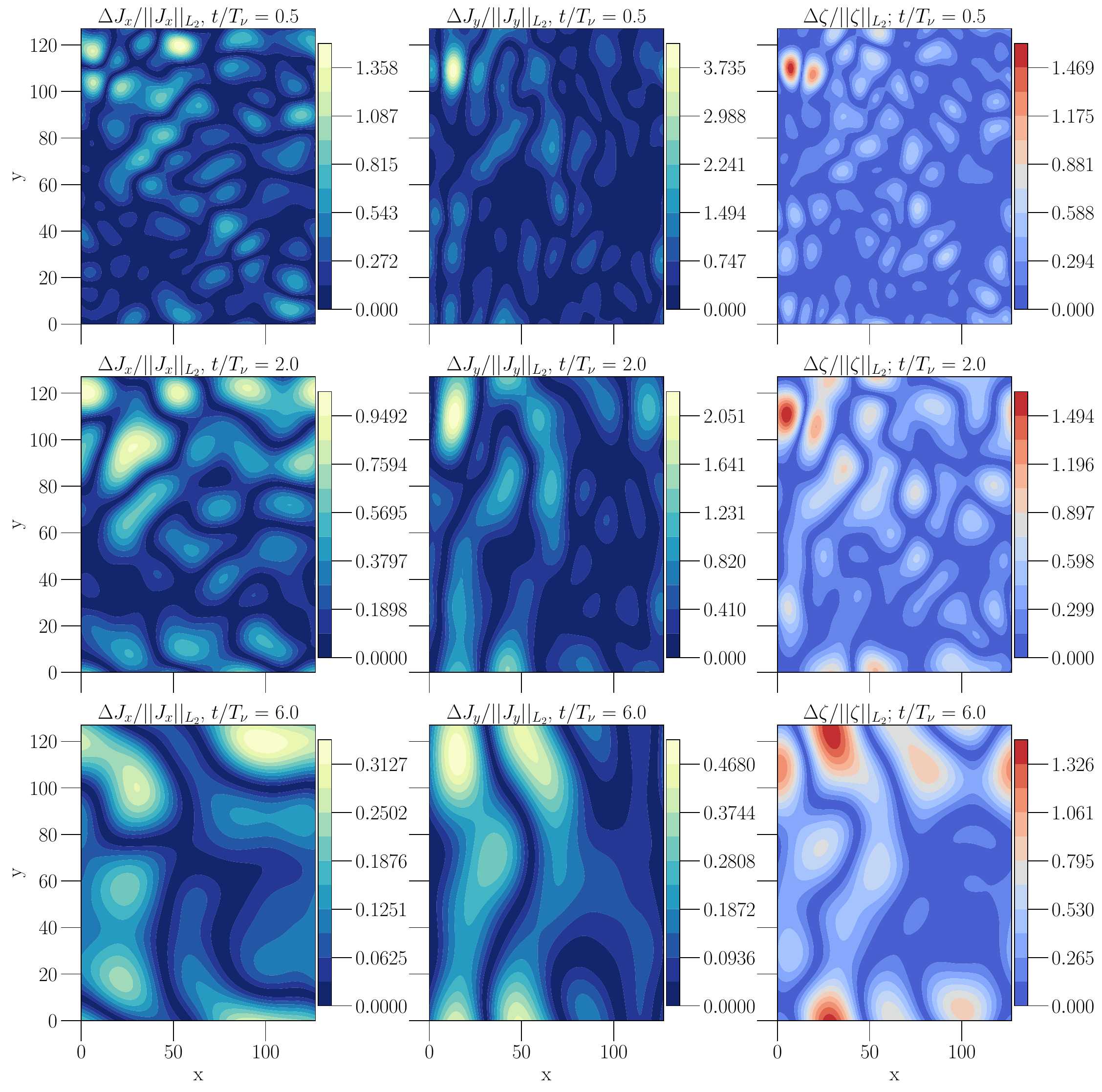}
    \caption{
    Evolution of the relative error in the two components of the momentum field $\mathbf{J}$ and in the vorticity $\zeta$ for an initial Reynolds number of approximately $46$ ($\omega=3/2$). 
    The relative error is evaluated according to Eq.~\eqref{eq:error} and time is normalized by the viscous timescale associated with the dominant mode of the initial condition, $T_\nu\approx467$ time steps. The time frames and figure layout are the same as in Fig.~\ref{fig:errors_re_15}, but the viscous timescale is three times longer.
    \\
    During the nonlinear transient, at $t/T_\nu=1/2$ and $t/T_\nu=2$, the local relative error reaches values of $\mathcal{O}(1)$, indicating large discrepancies in localized regions, particularly in the upper-left corner. Moreover, the error exceeds $10\%$ over large portions of the fields, even during the transient, indicating significant quantitative discrepancies between CLB2 and LBM.
    }
    \label{fig:error_re_46}
\end{figure}

In Fig.~\ref{fig:error_re_46}, we report the error fields for $J_x$, $J_y$, and $\zeta$, evaluated using Eq.~\eqref{eq:error}. As in the previous case, the errors are initially concentrated in the upper part of the domain. As the vortices coalesce, the regions with larger errors also tend to extend over larger areas. In this case, however, the error values are higher, suggesting a less accurate reproduction of the LBM solution, consistent with the visual comparison. Nevertheless, the global error appears to be lowest during the nonlinear transient, at $t/T_\nu=1/2$.

\begin{figure}
    \centering
    \includegraphics[width=0.75\linewidth]{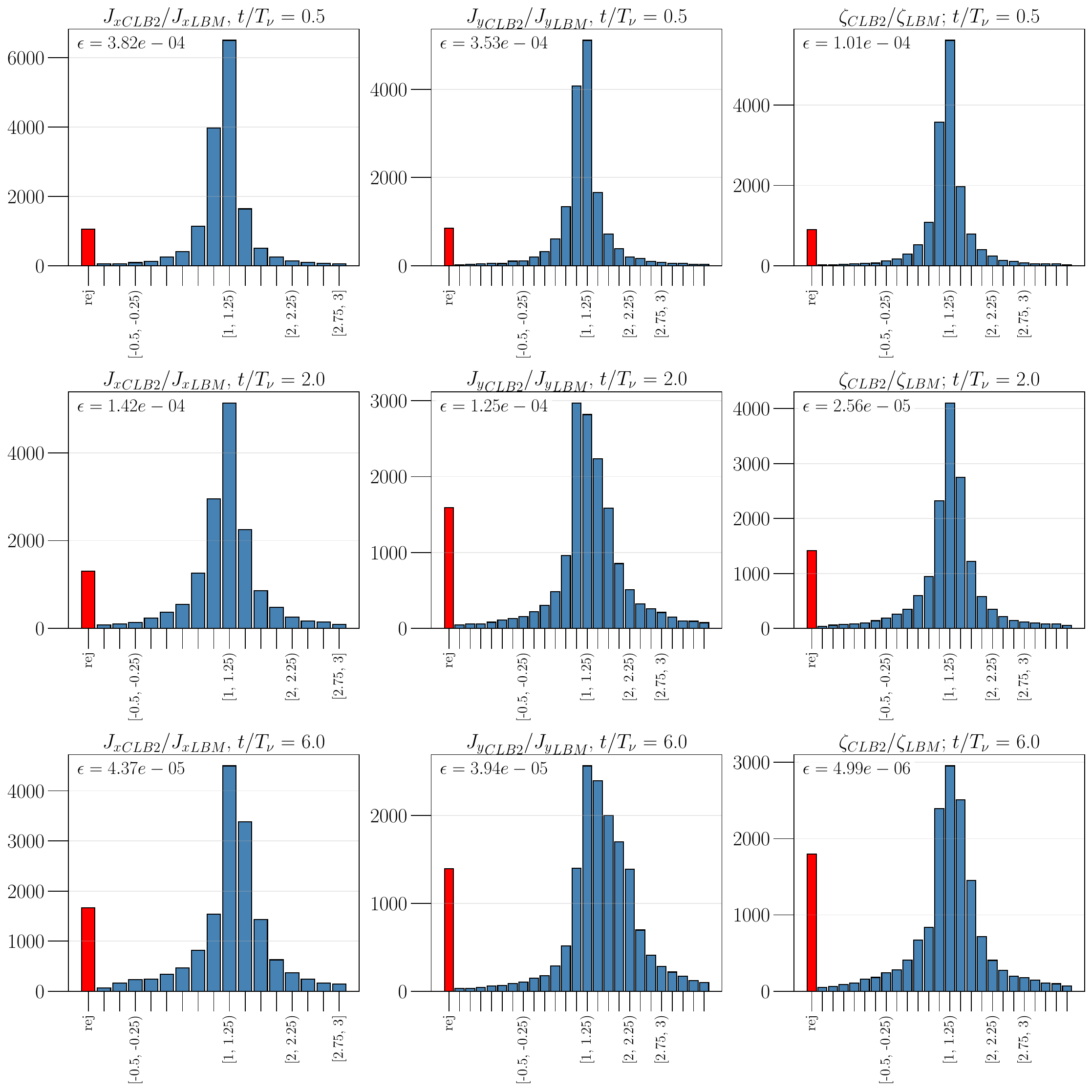}
    \caption{ 
   Histogram of the ratio between the fields obtained with LBM and CLB2 for the case shown in Fig.~\ref{fig:fields_re_46}, with an initial Reynolds number of approximately $46$. The figure layout is the same as in Fig.~\ref{fig:hsito_re_15}, with the threshold set according to Eq.~\eqref{eq:threshold}.
    \\
    The distribution remains mostly centered around $1$, but is broader than in the previous case; this indicates larger discrepancies between the two solutions that, however, do not significantly affect the overall field structure. As time progresses, the distribution broadens and the number of rejected values increases, indicating a gradual loss of agreement. This is particularly evident at $t/T_\nu\approx6$, when the number of rejected values becomes appreciable and the distribution is noticeably wider.
    }
    \label{fig:hist_re_46}
\end{figure}

To investigate whether these discrepancies arise from inaccuracies in the field amplitudes or from 
structural deformations, we examine the histograms of the ratios between CLB2 and LBM for the three fields, $J_x$, $J_y$, and $\zeta$. The results are shown in Fig.~\ref{fig:hist_re_46}. As in the previous case, the distributions remain centered around unity, with no clear evidence of a systematic scaling factor between the two solutions. However, the distributions are broader, and the number of outliers increases over time. Taken together, these observations suggest that CLB2 is still able to reproduce the overall field structure, although less accurately than in the case of $Re(0)\approx15$. The histograms also support the previous observation that the best agreement between the two methods occurs during the nonlinear transient.

\begin{figure}
    \centering
    \includegraphics[width=\linewidth]{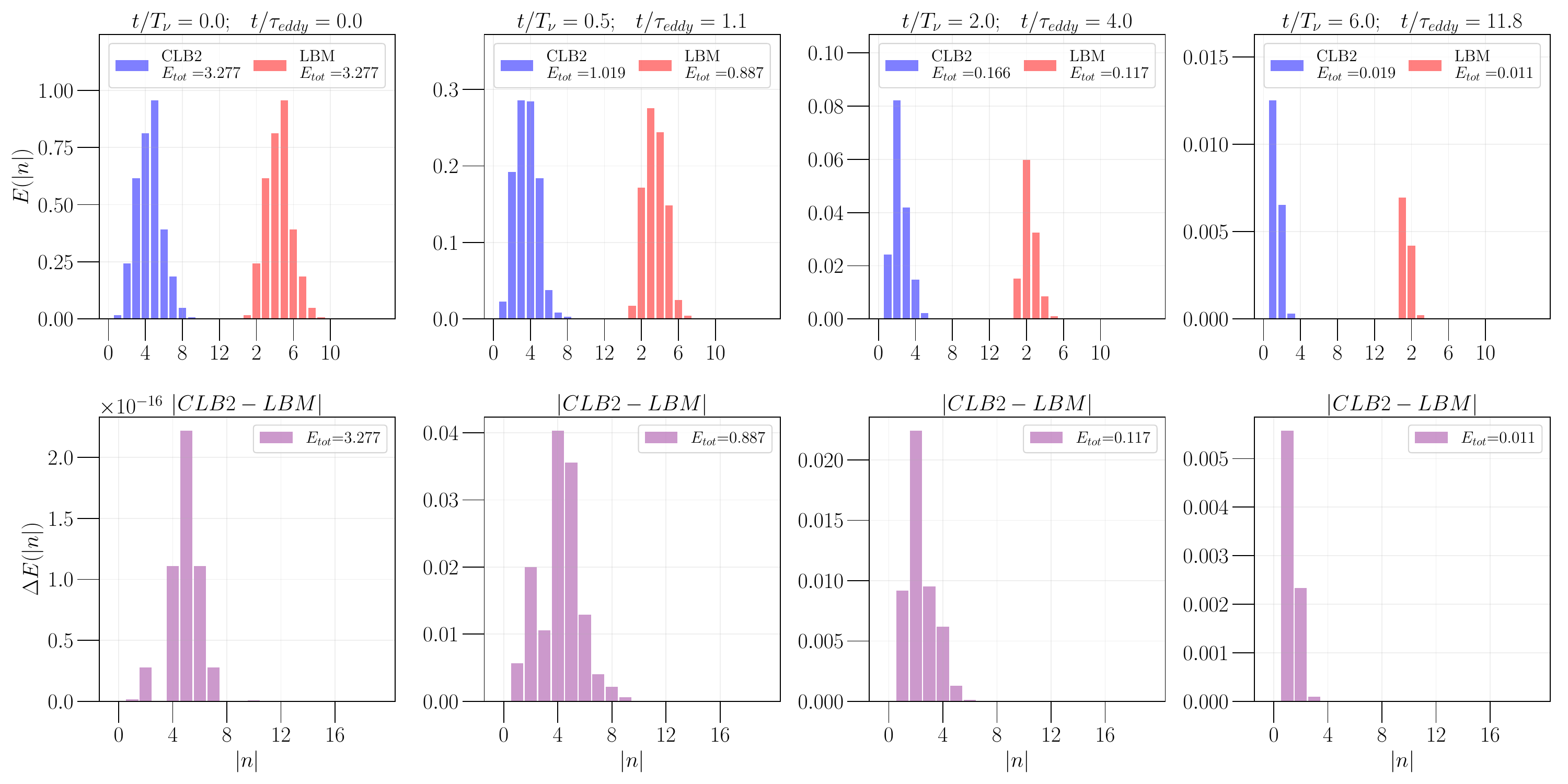}
    \caption{
    Evolution of the energy spectrum and the corresponding absolute error for CLB2 and LBM, with an initial Reynolds number of approximately $46$. The figure layout is the same as in Fig.~\ref{fig:energy_re_15} and refers to the configurations shown in Fig.~\ref{fig:fields_re_46}. The total energy obtained with each method is reported in the corresponding panel caption.
    \\
    As in the previous case, CLB2 correctly reproduces the energy spectrum without losing any modes. The only exception occurs at $t/T_\nu\approx0.5$, where the energy associated with the mode range $|n|\in[5,6)$ would be expected to be lower than that of the preceding mode. Overall, the energy distribution across the modes is well captured, while the main source of discrepancy appears to be the total energy, whote relative error exceeds $10\%$.
    }
    \label{fig:energy_re_46}
\end{figure}

We now turn to the energy spectrum to investigate how these discrepancies affect its evolution. Figure~\ref{fig:energy_re_46} shows the spectra obtained with CLB2 and LBM, together with the corresponding absolute error. The errors are larger than those observed at the lower Reynolds number. However, inspection of the spectra suggests that the main source of discrepancy is the higher energy values obtained with CLB2. This is also reflected in the relatively large errors in the total energy values reported in the panel captions.
\\
Despite these differences, the overall distribution of energy across the modes is still well reproduced. In particular, CLB2 retains all the modes present in the LBM spectrum throughout the evolution. The main discrepancy occurs during the transient, at $t/T_\nu=1/2$, when CLB2 gives an energy value for the mode $|n|=5$ that is close to that of the mode $|n|=4$, whereas in LBM the corresponding ratio is approximately $7/8$.
\\
Apart from this discrepancy, the evolution of the spectrum is consistent with the behavior observed with LBM. The number of visible modes progressively decreases from nine initially to eight, then five, and finally only three at $t/T_\nu\approx6$. Thus, although the energy values do not match perfectly, CLB2 correctly reproduces the progressive reduction in the number of visible modes.
\\
It is also worth noting that, at $t=2850$, the system has evolved for exactly six viscous times, as in the case with the lower initial Reynolds number. However, the corresponding number of eddy-turnover times, $\tau_{\mathrm{eddy}}$, is much larger, 
consistent with the lower dissipation, which allows the coherent structures to persist for longer. 

\begin{figure}
    \centering
    \includegraphics[width=0.85\linewidth]{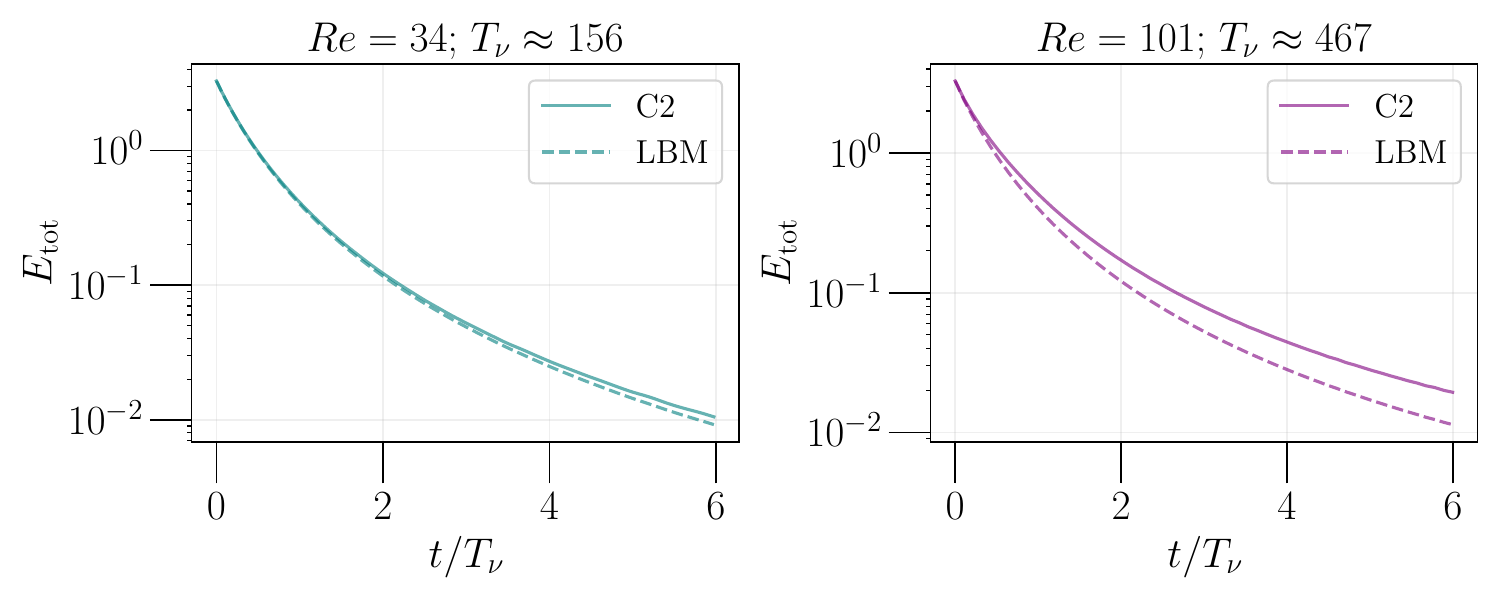}
    \caption{
    Total energy decay obtained with LBM and CLB2. Time is normalized by the viscous timescale $T_\nu$, as in the previous figures. 
    The evolution spans a total of 940 time steps for the lower Reynolds number and 2850 time steps for the higher Reynolds number. For $Re(0)\approx 46$, the energy decay is not accurately reproduced. Specifically, CLB2 yields higher total energy values than LBM.
    }
    \label{fig:placeholder}
\end{figure}

Since CLB2 consistently yields higher energy values than LBM in both cases, we further investigate the evolution of the total energy. Figure~\ref{fig:placeholder} shows the total energy decay for both Reynolds numbers. In both cases, the dynamics are dominated by viscous dissipation, which removes almost all the initial energy over the six viscous times considered. Nevertheless, CLB2 dissipates less energy than LBM, and this difference becomes more pronounced at the lower viscosity. Consequently, the discrepancies in the total energy become particularly large at late times in the higher-Reynolds-number case.

Overall, 
for an initial Reynolds number of $46$, CLB2 captures the main features of the turbulent-like decay without diverging from the reference evolution. However, it provides only a qualitative reproduction of the LBM solution, with significant discrepancies in field amplitudes and energy values.


\section{Summary}
In this paper, we have presented a quantitative analysis of the accuracy
of the second-order-truncated Carleman–Lattice Boltzmann method (CLB2) in
simulating  the nonlinear decay of a two-dimensional vortex flow at initial 
Reynolds numbers of $Re(0)\approx15$ and $Re(0)\approx46$ on a $128\times128$ grid. 

For this purpose, we developed a multi-GPU implementation to distribute the memory 
required by the second-order Carleman variables across multiple devices. 
To compare the two cases over equivalent timescales, we evolved both systems for 
six viscous times, corresponding to $940$ time steps for $Re(0)\approx 15$ 
and $2850$ time steps for the case with $Re(0)\approx46$, respectively.

In both cases, the initial vortex structures decay over time, while the smaller 
vortices merge into larger ones. This evolution is also reflected in the energy 
spectrum, where the number of visible modes progressively 
decreases as energy is transferred toward larger scales. 

At $Re(0)\approx15$, CLB2 correctly reproduces the nonlinear evolution of the 
flow, with local relative errors below $10\%$ in the momentum field 
$\mathbf{J}$ and the vorticity $\zeta$ over most  of the spatial domain. 
At $Re(0)\approx46$, however, CLB2 overestimates the amplitudes of these fields 
in certain regions. As the vortices coalesce, these regions tend to extend over larger 
areas, resulting in errors exceeding $10\%$ over most of the spatial domain. 
Nevertheless, the overall field structure is still captured, and the distribution of the 
CLB2/LBM ratios remains centered around unity even if more widely spread although it is broader than in the higher-viscosity case. 
This indicates that CLB2 preserves the main features of the flow, even though its accuracy degrades over time.
A similar behavior is observed in the energy spectra. At the lower Reynolds number, CLB2 reproduces the spectrum with high accuracy. At the higher Reynolds number, the overall distribution of energy across the visible modes is still captured, and no modes are lost. However, the energy values obtained with CLB2 are generally higher than those obtained with LBM. 
This discrepancy is also reflected in the total energy, which remains higher with CLB2 throughout the evolution. 


These results indicate  that CLB2 can sustain coherent structures over several
diffusive times, up to the point at which the flow has essentially decayed by several orders of magnitude.\
Several routes can potentially be pursued to mitigate the aforementioned loss of accuracy, such as
appropriate closures,  and regularization  techniques, and adaptive reference states~\cite{joseph25, REG, karlin2026residualcarlemanliftnonlinear}. 
Future work in this direction is clearly warranted and, in addition to further theoretical developments, will require the concurrent development of HPC-CLB codes to test these approaches on classical computers.

\section*{Acknowledgments}
The authors acknowledge valuable discussions with  M. Sanz, J.Gonzalez-Conde, D.Isla and C.Alvarez.
The authors have no conflicts to disclose. 
The data that support the findings of this study are available from the corresponding 
author upon reasonable request.

\bibliographystyle{unsrt}
\bibliography{biblio}

\end{document}